\documentclass[twocolumn, switch]{article}
\usepackage{preprint}
\usepackage{hyperref}
\usepackage{eso-pic}
\usepackage[numbers,square]{natbib}
\usepackage{graphicx}
\usepackage{multirow}
\usepackage{amsmath,amssymb,amsfonts}
\usepackage{amsthm}
\usepackage{mathrsfs}
\usepackage[title]{appendix}
\usepackage{xcolor}
\usepackage{textcomp}
\usepackage{booktabs}
\usepackage{algorithm}
\usepackage{algorithmicx}
\usepackage{algpseudocode}
\usepackage{listings}
\usepackage{caption}
\usepackage{float}
\usepackage{mathtools}
\hypersetup{colorlinks=true, linkcolor=purple, urlcolor=blue,
            citecolor=cyan, anchorcolor=black}
\usepackage[utf8]{inputenc}
\usepackage[T1]{fontenc}
\usepackage{lineno}
\usepackage{tikz}
\usetikzlibrary{spy,math,arrows}
\usepackage{newfloat}
\DeclareFloatingEnvironment[name={Supplementary Figure}]{suppfigure}
\usepackage{sidecap}
\sidecaptionvpos{figure}{c}
\newcommand{\R}{\mathbb{R}}

\newcommand{\D}{\mathcal{D}}
\newcommand{\Reg}{\mathcal{R}}
\newcommand{\Patch}{\mathbf{P}}

\newcommand{\defeq}{\vcentcolon=}
\definecolor{Periwinkle}{RGB}{123, 182, 220}
\theoremstyle{plain}

\theoremstyle{definition}

\usepackage{titlesec}
\titlespacing\section{0pt}{12pt plus 3pt minus 3pt}{1pt plus 1pt minus 1pt}
\titlespacing\subsection{0pt}{10pt plus 3pt minus 3pt}{1pt plus 1pt minus 1pt}
\titlespacing\subsubsection{0pt}{8pt plus 3pt minus 3pt}{1pt plus 1pt minus 1pt}
\definecolor{lime}{HTML}{A6CE39}
\DeclareRobustCommand{\orcidicon}{
	\begin{tikzpicture}
	\draw[lime, fill=lime] (0,0) 
	circle [radius=0.16] 
	node[white] {{\fontfamily{qag}\selectfont \tiny ID}};
	\draw[white, fill=white] (-0.0625,0.095) 
	circle [radius=0.007];
	\end{tikzpicture}
	\hspace{-2mm}
}
\foreach \x in {A, ..., Z}{\expandafter\xdef\csname orcid\x\endcsname{\noexpand\href{https://orcid.org/\csname orcidauthor\x\endcsname}
			{\noexpand\orcidicon}}
}

\title{Semantic Segmentation for Histopathology using Learned Regularization based on Global Proportions}
\AddToShipoutPictureBG*{%
  \AtPageLowerLeft{%
    \raisebox{0.5\paperheight}{%
      \hspace{0.02\paperwidth}%
      \rotatebox{90}{\scalebox{1.2}{%
        \href{https://doi.org/}{\textcolor{gray}{Publication doi}}}}}}}
\AddToShipoutPictureBG*{%
  \AtPageLowerLeft{\hspace{0.12\paperwidth}%
    \raisebox{0.04\paperheight}{\scalebox{0.8}{%
     \textcolor{gray}{\shortstack[l]{%
        $^{*}$ and $^{\dagger}$ contributed equally to this work.\\
        \P\,correspondence: \texttt{effland@iam.uni-bonn.de}}}}}}}
\usepackage{authblk}

\author[1\thanks{\tt{s59yli@uni-bonn.de}, equal contribution.}]{Yangping Li\orcidA{}}
\author[2\thanks{\tt{pinetz@iam.uni-bonn.de}, equal contribution.}]{Thomas Pinetz\orcidB{}}
\author[3\thanks{\tt{Michael.Hoelzel@ukbonn.de}}]{Michael H\"{o}lzel\orcidC{}}
\author[1\thanks{\tt{Marieta.Toma@ukbonn.de}}]{Marieta Toma}
\author[2\thanks{\tt{effland@iam.uni-bonn.de}}]{Alexander Effland\orcidE{}}

\affil[1]{Institute of Pathology, University Hospital Bonn, Sigmund-Freud-Str.\ 25, 53127 Bonn, Germany}
\affil[2]{Institute for Applied Mathematics, University of Bonn, Endenicher Allee 60, 53115 Bonn, Germany}
\affil[3]{Institute of Experimental Oncology, University Hospital Bonn, Sigmund-Freud-Str.\ 25, 53127 Bonn, Germany}

\begin{document}

\twocolumn[\begin{@twocolumnfalse}
\maketitle 

\begin{abstract}
In pathology, the spatial distribution and proportions of tissue types are key indicators of disease progression, and are more readily available than fine-grained annotations.
However, these assessments are rarely mapped to pixel-wise segmentation.
The task is fundamentally underdetermined, as many spatially distinct segmentations can satisfy the same global proportions in the absence of pixel-wise constraints.
To address this, we introduce Variational Segmentation from Label Proportions (VSLP), a two-stage framework that infers dense segmentations from global label proportions, without any pixel-level annotations.
This framework first leverages a pre-trained transformer model with test-time augmentation to produce a pixel-wise confidence estimate.
In the second stage, these estimates are fused by solving a variational optimization problem that incorporates a Wasserstein data fidelity term alongside a learned regularizer.
Unlike end-to-end networks, our variational method can visualize the fidelity--regularization energy, resulting in more interpretable segmentation.
We validate our approach on two public datasets, achieving superior performance over existing weakly supervised and unsupervised methods.
For one of these datasets, proportions have been estimated by an experienced pathologist to provide a realistic benchmark to the community.
Furthermore, the method scales to an in-house dataset with noisy pathologist labels, severely outperforming state-of-the-art methods, thereby demonstrating practical applicability.
The code and data will be made publicly available upon acceptance at \url{https://github.com/xiaoliangpi/VSLP}.\\
\end{abstract}


\vspace{0.5cm}
\end{@twocolumnfalse}]


\section{Introduction}\label{sec1}

Proportions of distinct tissue patterns, estimated by pathologists, serve as sensitive markers for monitoring disease progression and evaluating treatment efficacy, particularly in cancer patients~\cite{chen24mul,jas25his}.
Pathologists are trained to estimate these ratios accurately, often building extensive datasets of such assessments.
However, automated models trained on these datasets typically predict only global ratios and operate as ``black boxes,'' providing no insight into the spatial distribution of tissue types.
To provide the transparency and interpretability required by clinicians, global estimates should be visualized as semantic segmentation maps that reveal both the extent and location of each tissue class.
Inferring pixel-wise segmentation from slide-level proportions is challenging due to the gigapixel resolution and heterogeneity of whole-slide images~\cite{levy2020spatial}, which makes substantial downscaling infeasible without losing critical detail.
Moreover, a single global proportion can correspond to an exponential number of possible pixel-wise layouts, rendering the problem fundamentally underdetermined.

The task of inferring pixel-wise segmentation from global tissue proportions can be formulated as a \textit{learning from label proportions} (LLP) problem~\cite{arde17co}.
In this context, a whole-slide image is treated as a bag~\cite{okuo2023proportion}, with each pixel representing an instance, while labels are provided as proportions of instances at the bag level.
The primary goal in LLP is therefore to train an instance classifier using only these bag-level label proportions.
Existing approaches struggle to resolve this ambiguity.
Class Activation Maps~\cite{ChHo19} produce sparse and fragmented masks without utilizing precise proportion information.
Multiple instance learning~\cite{li22onl,fang23weak} methods typically process patches independently, thereby ignoring global consistency constraints.
Scribble-based supervision~\cite{liu22scribble} adds labeling burden, ignoring the global proportion information already available from pathologists.
Crucially, these approaches offer limited capacity to resolve ambiguity among multiple plausible segmentations or to quantify posterior uncertainty.

Several prior works mitigate global-to-local ambiguity in weakly supervised segmentation by imposing additional constraints.
Semantic affinity--based methods propagate sparse cues, such as class activation maps.
These cues are used to generate spatially coherent pseudo-masks~\cite{18irn}.
More recent approaches incorporate uncertainty estimation to suppress noisy pseudo-labels during training~\cite{li22urn,24flipcam}.
However, these methods still produce only a single segmentation and do not represent multiple configurations consistent with the same global supervision.
Graph-based formulations also enforce global constraints.
However, they do not characterize uncertainty over segmentation layouts~\cite{zh21joint}.
Beyond weak supervision, global--local segmentation has been studied in fully supervised settings.
These methods fuse global context with local features~\cite{lin20glnet}.
They assume access to pixel-level annotations.
In contrast, in the LLP scenario explored here, the key challenge lies in resolving inference ambiguity arising from global-only supervision.

\begin{figure*}[ht]
    \centering
    \includegraphics[width=.95\linewidth]{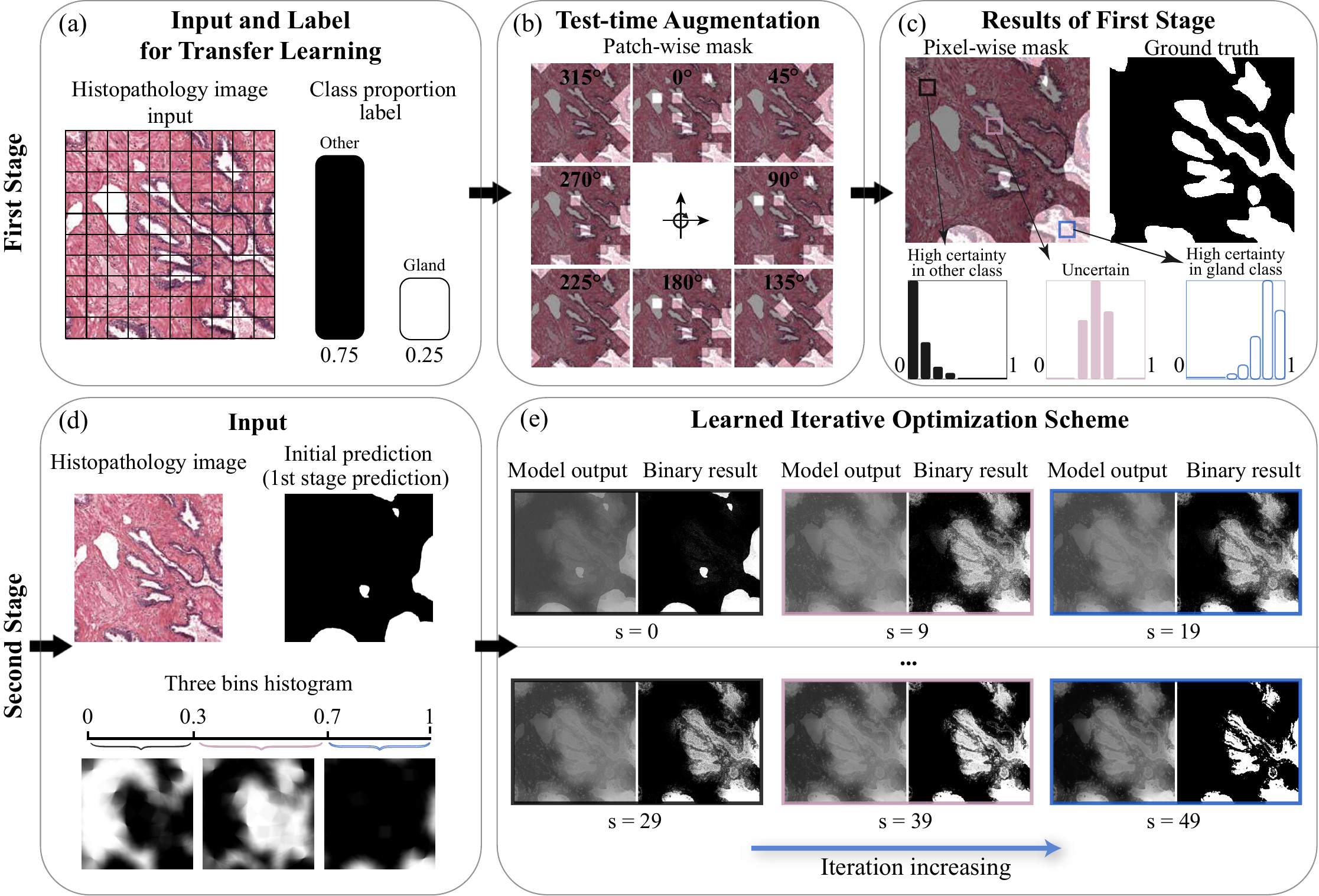}
    \caption{Semantic segmentation masks are generated by VSLP for histopathology images using a two-stage approach. Initially, (a) an ImageNet~\cite{DeDo09} pre-trained Swin Transformer~\cite{LiLi21} is fine-tuned on global class proportion labels by averaging the predictions across image patches. In (b), test-time augmentation is applied to generate predictions for various rotations, which are subsequently warped back to the original orientation. These proportions are pooled in pixel-wise histograms as well as the initial predictions by thresholding (c,d). A variational refinement problem is solved using a learned iterative optimization scheme in the second stage. The output thereof across multiple iterations is illustrated in (e).}
    \label{fig:workflow}
\end{figure*}

Addressing such ambiguity requires incorporating prior structure in addition to data-driven cues.
Variational methods provide a principled framework for this purpose and have long been used to address inverse problems such as image reconstruction~\cite{RuOs92,liu15unified}.
The central idea is to determine the maximum a posteriori estimator by optimizing an energy function composed of a data fidelity term and a regularizer~\cite{ChPo16}.
From a Bayesian perspective, the fidelity term reflects how well a segmentation agrees with observed data.
While classical approaches often use handcrafted priors~\cite{ChPo16}, these can be too simplistic to capture the rich, heterogeneous textures found in histopathology.
Recent advances demonstrate that learned, non-convex regularizers can model these complex patterns more effectively~\cite{KoEf21,GoNe24}.
Moreover, model-based methods can be more sample-efficient and scalable than large end-to-end networks~\cite{KoEf20,PiKo22}, making them well-suited for high-resolution images.

In this work, we propose a two-step framework termed \textbf{V}ariational \textbf{S}egmentation from \textbf{L}abel \textbf{P}roportions (VSLP) that transforms slide tissue proportions into dense, pixel-wise segmentations without detailed annotations.
In the first stage, we leverage test-time augmentation (TTA) to sample multiple segmentation outputs under varying input perturbations, given only the global proportion.
Inspired by the work of~\cite{TaPe23}, the TTA output is treated as samples from an unknown probability distribution, yielding an approximate posterior via pixel-wise histograms.
In the second step, we refine these coarse distributions by minimizing a variational energy functional, combining a Wasserstein-based data fidelity term (enforcing consistency with the first stage) and a learned regularizer that captures histological priors.
This optimization is performed using an iterative scheme based on the Unadjusted Langevin Algorithm (ULA)~\cite{Dala16}, and incorporates diffusion-inspired updates and a GMM variant for uncertainty modeling.
The result is a spatially coherent, uncertainty-aware segmentation (see Figure~\ref{fig:workflow}).
The main contributions of this paper are:
\begin{itemize}
  \item We introduce a variational, uncertainty-aware framework for resolving global-to-local ambiguity in segmentation from label proportions.
  \item We align TTA-derived posteriors with learned priors via variational inference, producing smooth, spatially coherent, and interpretable segmentations.
  \item We leverage the stochasticity in ULA, GMM, and diffusion to capture posterior variability, adaptively refine uncertain regions, and preserve reliable structures.
  \item We demonstrate state-of-the-art performance on two public datasets and strong generalization on an in-house dataset with noisy clinical labels.
\end{itemize}

\section{Related Work}\label{sec2}
In this section, we review four key areas relevant to our work: weakly-supervised methods, including Multiple Instance Learning (MIL), Learning from Label Proportions (LLP), and variational methods that incorporate regularization to handle uncertainty.
We highlight how their integration motivates our proposed method.

\subsection{Weakly-Supervised Semantic Segmentation}\label{subsec2_1}
Weakly supervised semantic segmentation aims to generate fine-grained segmentation masks from coarse-grained annotations, significantly reducing the need for labor-intensive pixel-level labeling~\cite{zh21joint}.
This approach involves different levels of labeling granularity, including image- or patch-level labels~\cite{liu23tssk}, bounding boxes~\cite{liang21}, scribbles~\cite{liu22scribble}, and points~\cite{zhang23weak}.
Among these, image-level labeling is the least costly but presents unique challenges in histopathology due to high inter-class similarity and indistinct boundaries~\cite{chan21compre}.
Historically, weakly-supervised approaches leveraging class labels have primarily relied on Multiple-Instance Learning (MIL).

\subsection{Multiple-Instance Learning}
Previous MIL approaches in medical imaging primarily focused on bag-level classification, where images or patches are treated as bags and pixels as instances~\cite{shar21c}.
Consequently, weakly supervised segmentation is reformulated as an instance prediction task based on bag-level annotations.
In MIL, a bag is labeled as positive if at least one instance within it is positive~\cite{fang23weak}.
However, this property often results in gradient updates focusing on significant positive instances, while other instances suffer from vanishing gradient issues~\cite{das20detec}, leading to incomplete segmentation.
Despite this limitation, MIL has been employed with class activation maps for semantic segmentation~\cite{ChHo19,han22mul,li22onl}.
Some MIL-based methods improve pseudo-label quality by uncertainty-aware reweighting~\cite{li22urn}.
In contrast to our work, MIL methods~\cite{li23weak} rely on local (bag-level) supervision, whereas we use \emph{global} image-level label proportions.

\subsection{Learning from Label Proportions}
LLP builds instance classifiers using only aggregated label proportions from bags or patches, relying on global information~\cite{arde17co}.
Early methods focused on single-category proportions~\cite{yu13svm}, but later evolved to handle multi-category tasks by averaging predicted class probabilities across instances~\cite{yang21two}.
Most LLP methods employ a proportionality loss between true proportions and mean predictions~\cite{mat23learn}.
Even bag-level annotations are often insufficient for training accurate models, and existing studies have investigated techniques such as data augmentation~\cite{tang24all} and pseudo-labeling~\cite{tok20neg} to improve performance.
Related weakly supervised methods further refine predictions through TTA~\cite{24flipcam}.
In histopathology, FGNet introduces a graph-based LLP formulation to enforce global proportion consistency~\cite{zh21joint}.
Recently, \cite{MaSu24} proposed the first LLP-related approach in histopathology by considering partial labels.
Using local (patch) binary labels, MIL is first trained to differentiate positive and negative instances.
For positive cases, category proportions are estimated from patch classification results, effectively combining MIL with LLP.
In contrast, we treat the task as a classical LLP problem and use a \emph{single} global proportion per image.

\subsection{Variational Methods}\label{subsec2_2}
Historically, image reconstruction tasks such as denoising and deblurring have been formulated as variational problems solved via iterative optimization schemes~\cite{RuOs92,ChPo11}.
These methods minimize an energy composed of a \emph{data fidelity term}, measuring consistency with observed data, and a \emph{regularizer} encoding natural image statistics~\cite{ChPo16}.
While the optimal data fidelity term is well understood for certain problems (e.g., Gaussian denoising), it remains unclear for more general settings~\cite{PiKo21}.
In some cases, data fidelity is instead enforced through data-consistency layers embedded within network architectures~\cite{ZhGo20,XiDo21}.

Traditionally, regularizers were handcrafted to promote common image structures, such as total generalized variation (TGV)~\cite{BrKu10}, but these designs often fail to capture the complexity of real-world data~\cite{KoEf21}.
Recently, data-driven regularizers based on deep learning have shown superior performance by modeling richer image priors~\cite{lun22learn,GoNe24}.
Such iterative schemes constrain network outputs toward physically consistent solutions, improving generalization with fewer parameters~\cite{PiKo22}.
Variational formulations have also been widely applied to image segmentation, where segmentation is obtained by minimizing region- or boundary-based energy functionals~\cite{shu25variat}.
These approaches often employ level-set representations and shape priors to resolve ambiguous local evidence~\cite{liu15unified}.
In the context of LLPs, prediction uncertainty arises from noisy and ambiguous labels; regularization mitigates this issue by incorporating prior knowledge, as confirmed by our results.

\section{Method}\label{sec3}
Our method consists of two stages as visualized in Figure~\ref{fig:workflow}.
First (see Section~\ref{subsec3_1}), a Swin Transformer~\cite{LiLi21} is fine-tuned to provide an accurate representation of the proportions present in the image.
Then, TTA is used to produce a pixel-wise distribution, inspired by~\cite{TaPe23}.
In a second step (see Section~\ref{subsec3_2}), this distribution is fused to generate a final segmentation mask by solving a variational problem involving learned regularizers~\cite{KoEf20}.

\subsection{First Stage}\label{subsec3_1}
We start with a dataset of $N$ high-resolution histopathology images $x_i\in\R^{H\times W\times C}$, where $H$ and $W$ denote the height and width of the images, and $C$ is the number of color channels.
Each image is associated with a label proportion vector $y_i$ over $N_y$ semantic classes, such that $y_i$ lies in the $N_y$-dimensional probability simplex.

Due to the large size of whole-slide images, we divide each image into $N_p$ overlapping patches using a patching function $\Patch: \R^{H\times W \times C} \to \R^{N_p\times H_p\times W_p\times C}$, where $H_p$ and $W_p$ denote the patch size.
Overlapping patches help reduce boundary artifacts.
To estimate class proportions locally, we apply a Swin Transformer~\cite{LiLi21}, pre-trained on ImageNet~\cite{DeDo09}, to each patch.
This model $f:\R^{H_p\times W_p\times C}\to\R^{N_y}$ outputs a class proportion vector for each patch.
Since patches may vary in tissue content, we weight each patch prediction by a factor $\epsilon_j$, representing the fraction of tissue in patch $j$ relative to the total tissue in the slide.
The final prediction for the image is obtained by averaging all patch predictions:
\begin{equation}\label{eq1}
    \ell(x, y) = \frac 1 N \sum^N_{i=1}\left\Vert y - \frac 1 {N_p} \sum^{N_p}_{j=1} \epsilon_j f((\Patch x_i)_j)\right\Vert^2_{2}.
\end{equation}
This approach enables accurate learning of patch-level predictions from global label proportions.
However, it does not provide pixel-wise segmentation, and spatial variation remains limited to per-patch class distributions.

To achieve pixel-wise predictions, we apply a simple yet effective TTA~\cite{KiKi20,ShBl21}: we rotate the image $N_r$ times by evenly spaced angles between $[0^\circ, 359^\circ]$.
For each rotation, we generate patch-wise predictions, which are then rotated back to the original orientation.
We obtain $N_r$ predictions per pixel, denoted as $z_i\in \R^{N_r\times H\times W\times N_y}$, as illustrated in Figure~\ref{fig:workflow}(b).

Next, we generate a semantic segmentation by selecting the class with the maximum number of votes for each pixel, denoted as $\bar{x}$ and visualized in Figure~\ref{fig:workflow}(c) using alpha blending.
Although this approach produces interpretable pixel-wise segmentation masks, it discards valuable information regarding the confidence margins between competing classes, making it less effective for small objects or sharp boundaries.

\subsection{Second Stage}\label{subsec3_2}
The first stage provides multiple predictions per pixel through rotation-based augmentation, but the simple voting mechanism discards valuable uncertainty information.
To better leverage this rich information while producing spatially coherent segmentation, we formulate the problem as a variational optimization framework.
Simply put, our approach seeks to find an optimal segmentation $u_i^*$ that balances two objectives:
A data fidelity term $\mathcal{D}(u, z_i^h)$ ensures that the solution remains consistent with the prediction evidence from the first stage.
In addition, a regularization term $\mathcal{R}$ promotes spatially coherent and structurally reasonable segmentation by leveraging learned priors.
In summary, this approach leads to the variational problem:
\begin{equation}\label{eq2}
u_i^* = \arg\min_u \mathcal{D}(u, z_i^h) + \lambda \mathcal{R}(u; z_i^h, x_i, \theta^*),
\end{equation}
where $z_i^h$ denotes the uncertainty information from stage one, $x_i$ is the histopathology image, $\theta^*$ are parameters of the learned prior, and $\lambda$ balances the two terms.

\subsubsection{Data Fidelity Term}
Rather than treating the $N_r$ predictions per pixel as independent votes, we adopt a statistical perspective inspired by Tachella et al.~\cite{TaPe23}, where each pixel is modeled as a probability distribution over the $N_y$ classes.
This allows us to capture the full uncertainty information present in our multi-prediction setup.

We convert the discrete predictions $z_i \in \mathbb{R}^{N_r \times H \times W \times N_y}$ into continuous probability distributions by using $N_h$ bins with equidistant centers $b_l$, $l = 1, \dots, N_h$, in the interval $\left[\frac{1}{2N_h}, 1 - \frac{1}{2N_h}\right]$, computing histograms $z_i^h \in \mathbb{R}^{N_h \times H \times W \times N_y}$.
Our target segmentation $u_i$ is treated as a pixel-wise $\delta$-distribution centered at the predicted class intensities.
To measure the deviation between these distributions, we employ the Wasserstein distance~\cite{AbGi05,Vi09}.
It offers several advantages over alternatives like the Kullback--Leibler divergence or Jensen--Shannon distance~\cite{NoCs16}:
It is defined even when probability distributions do not overlap and provides informative gradients for optimization~\cite{ArCh17}.
For our $\delta$-distribution setup, this simplifies to the following closed-form expression:
\begin{equation}\label{eq3}
\mathcal{D}(u_i, z_i^h) \defeq \frac{1}{2} \mathcal{W}_2^2(u_i, z_i^h) = \frac{1}{2} \sum_{h=1}^H \sum_{w=1}^W \sum_{k=1}^{N_y} \sum_{l=1}^{N_h} (u_{i,h,w,k} - b_l)^2 z_{i,l,h,w,k}^h.
\end{equation}

\subsubsection{Learned Regularization Term}
While the Wasserstein distance effectively aligns our segmentation with the first-stage predictions, it does not incorporate the underlying image content or account for the spatial structure of uncertainty.
In particular, it cannot model the relationship between texture patterns in the histopathology image $x_i$ and the uncertainty encoded in $z_i^h$.
To address this, we introduce a learned regularizer $\mathcal{R}(u; z_i^h, x_i, \theta^*)$, implemented as a convolutional neural network, following~\cite{PiKo21,KoEf21}.
Critically, we incorporate both the original image $x_i$ and the pixel-wise histogram $z_i^h$ as additional input channels.

To implement this regularizer, we design a U-Net variant~\cite{ron15u}.
The model follows an encoder-decoder structure.
The encoder extracts hierarchical features through convolutional layers with progressively increasing channels, while the decoder refines spatial details via transposed convolutions and skip connections.
The network parameters $\theta^*$ are learned during training to capture the relationship between image features and the optimal segmentation refinement.

\subsubsection{Optimization}
To solve the variational optimization problem defined in Equation~\eqref{eq2}, we employ the ULA~\cite{11bayes}, a gradient-based iterative method combined with stochastic perturbations.
This approach enables our segmentation model to explore multiple plausible solutions and avoid suboptimal local minima by injecting Gaussian noise during optimization.

Starting from the coarse segmentation $\bar{x}$ obtained in the first stage, ULA updates the segmentation iteratively according to:
\begin{equation}\label{eq4}
u_i^{s+1} = u_i^s - \frac{\tau}{S}\left(\nabla_u \mathcal{D}(u_i^s,z_i^h)+\lambda\nabla_u\mathcal{R}(u_i^s;z_i^h,x_i,\theta^*)\right) + \sqrt{2\frac{\tau}{S}}\,\mathbf{n},
\end{equation}
where $s=1,\dots,S$ denotes iteration steps, $\tau>0$ is a learnable step-size parameter, and $\mathbf{n}\sim\mathcal{N}(0,I)$ introduces randomness to explore segmentation uncertainty.

\begin{figure}[htb]
    \centering
    \includegraphics[width=.99\linewidth]{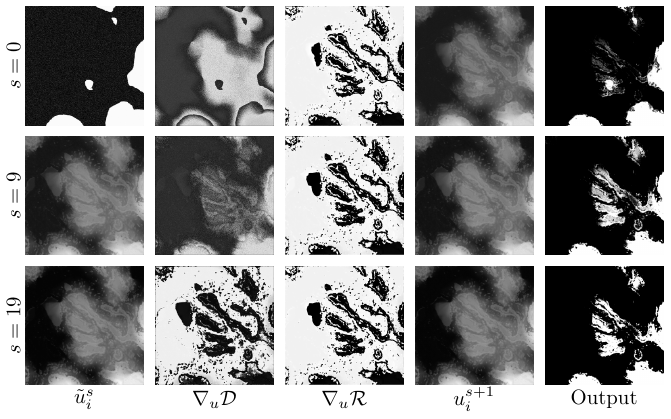}
    \caption{Visualization of the iterative optimization scheme for various $s=\{0, 9, 19\}$ (trained for $S=20$), showing the current refinement as well as the two contributing gradients, i.e., data fidelity term $\D$ and regularizer $\Reg$. The scheme is defined in \eqref{eq5}. The final output is obtained by thresholding.}
    \label{fig:iterations}
\end{figure}

Inspired by denoising diffusion models~\cite{nic2021}, we further enhance ULA with a noise scheduling strategy.
Specifically, we replace the fixed noise level with a linearly decreasing schedule:
\begin{equation}\label{eq5}
u_i^{s+1} = u_i^s - \frac{\tau}{S}\left(\nabla_u\mathcal{D}(u_i^s,z_i^h)+\lambda\nabla_u\mathcal{R}(u_i^s;z_i^h,x_i,\theta^*)\right)+\left(1-\frac{s}{S}\right)\sigma^0\,\mathbf{n}.
\end{equation}
This scheme ensures broad exploration in early iterations, progressively focusing on detailed refinements in later iterations.
This iterative refinement process, along with the input channels comprising the original image $x_i$, initial segmentation $\bar{x}$, and histogram $z_i^h$, is visually summarized in Figure~\ref{fig:workflow}(d).
We refer to the final segmentation $u_i^S$ obtained after these iterations as VSLP$_{Diff.}$, emphasizing its use of Wasserstein-based data fidelity, learned spatial priors, and Langevin-based refinement.
The overall refinement process is illustrated in Figure~\ref{fig:iterations}.

Despite the effectiveness of ULA and diffusion-based methods, they rely on high-dimensional histograms $z_i^h\in\R^{N_h\times H\times W\times N_y}$, and random noise injections can be computationally burdensome.
To improve computational efficiency and obtain explicit uncertainty estimates, we introduce a Gaussian Mixture Model (GMM) variant, which replaces the high-dimensional histogram representations with an easily computable Gaussian parameterization.
Specifically, instead of representing uncertainty in an $N_h$-bin histogram $z_i^h$, we fit a single Gaussian with mean $\mu_i$ and standard deviation $\sigma_i$, reducing the representation to $\R^{2\times H\times W\times N_y}$.
Note that this approach can be easily extended to multi-component GMMs.
We compute the Gaussian parameters by the method of moments, i.e., for given bin centers $b_l$, $l = 1, \dots, N_h$ and histogram weights $z_i^h \in \mathbb{R}^{N_h \times H \times W \times N_y}$ we obtain
\begin{equation}\label{eq6}
(\mu_i^0,\sigma_i^0)=\left(\frac{\sum_{l=1}^{N_h}z_i^h b_l}
      {\sum_{l=1}^{N_h}z_i^h},
\max\left\{\sqrt{\frac{\sum_{l=1}^{N_h}z_i^h(\mu_i^0-b_l)^2}
      {\sum_{l=1}^{N_h}z_i^h}},10^{-6}\right\}\right),
\end{equation}
where the clamping of the standard deviation is performed for stability reasons.
We embed the GMM representation into our variational framework by redefining the data fidelity and regularization terms to operate on $(\mu_i,\sigma_i)\in\R^{2\times H\times W\times N_y}$.
The refinement step of \eqref{eq2} thus becomes (for simplicity, we do not change the notation of $\mathcal{D}$ and $\mathcal{R}$)
\begin{align}\label{eq7}
(\mu_i^{s+1},\sigma_i^{s+1}) =(\mu_i^{s},\sigma_i^{s})-\frac{\tau}{S}\Big(&\nabla_{(\mu_i^s,\sigma_i^s)} \mathcal{D}((\mu_i^{s},\sigma_i^{s}),(\mu_i^0,\sigma_i^0))\notag\\
&+\lambda\nabla_{(\mu_i^s,\sigma_i^s)}\mathcal{R}((\mu_i^{s},\sigma_i^{s});(\mu_i^0,\sigma_i^0),x_i,\theta^*)\Big).
\end{align}
Notably, the regularizer can now utilize the uncertainty information encoded in $\sigma_i$ to apply adaptive smoothing.

Finally, parameters $(\theta,\lambda,\tau)$ are optimized on the training set by minimizing the mean squared error (MSE) between segmentation predictions and known class proportions:
\begin{equation}\label{eq8}
\arg\min_{(\theta,\lambda,\tau)}\sum_{i=1}^N\left\Vert\frac{1}{HW}\sum_{h=1}^{H}\sum_{w=1}^{W}u^S_{i,h,w}-y_i\right\Vert_2^2.
\end{equation}

\section{Numerical Results}\label{sec4}

In this section, we present the performance of our proposed method on two benchmark datasets and one in-house dataset, comparing it against several state-of-the-art approaches.
Additionally, we conduct a series of ablation studies to quantitatively assess the sensitivity of the method to various factors.

\subsection{Experimental Setup}\label{subsec4_1}

In what follows, we present the details of the datasets and implementation settings.
Upon acceptance, the code will be made available to easily reproduce our results.

\paragraph{Datasets.}
We evaluate our method on two publicly available histopathology datasets and one in-house dataset.
A toy example illustrating our setup is provided in the supplementary materials.
Henceforth, results from our ULA, GMM, and diffusion variants are denoted as VSLP$_{ULA}$, VSLP$_{GMM}$, and VSLP$_{Diff.}$, respectively.

The RINGS dataset~\cite{salvi21hybrid} comprises 1500 H\&E-stained slides from 150 patients, each measuring $1500\times1500$ pixels and acquired at $100\times$ magnification.
Expert pathologists annotated two categories: non-glandular tissue (category~0) and prostate glands (category~1).
In addition to pixel-derived slide-level proportions, we collected coarse proportion estimates provided by pathologists to reflect realistic clinical conditions, which will be released as an additional benchmark dataset.

The BCSS dataset~\cite{amgad2019struc} contains 151 histopathology slides from The Cancer Genome Atlas Program (TCGA).
Ground truth annotations use pixel values from 0 to 21, representing 22 tissue categories; pixel value~1 denotes tumor regions, while others correspond to non-tumor tissues.
To ensure each category exceeded a proportion of 30\%, we grouped images into two classes: 0 (other tissues) and 1 (tumors).
Due to the high resolution of each slide (typically $>2000\times2000$ pixels), we applied a sliding window with a step size of 1500 pixels, discarding excess pixels.
Both the RINGS and BCSS datasets were randomly split into train/validation/test sets in an $8\colon 1 \colon 1$ ratio.

To validate the segmentation performance on datasets, we employed three evaluation metrics: Dice coefficient, mean Intersection over Union (mIoU), and $95\%$ Hausdorff Distance (95HD).
All metrics were computed over the entire image rather than on individual patches.

The in-house dataset was collected at the University Hospital Bonn and comprises 325 histopathology images of renal cell carcinoma stained with CD31 and scanned at $10\times$ magnification.
For evaluation, $80\%$ of the images were used for training and $20\%$ for testing.
A board-certified pathologist with over 20 years of experience annotated each slide with global label proportions of three vascular patterns: high branching (145 images), low branching (97 images), and sinusoid (83 images).
Each image was analyzed individually, with the predominant pattern defining the ground truth class.
Annotations focus exclusively on vascular architecture, with class proportions defined within the vascular compartment.
Tumor cells and background regions are excluded to emphasize angiogenesis-related patterns.
In the absence of ground truth segmentation, we report Accuracy, Precision, Recall, and F1 scores for the predominant class, as well as MAE and MSE for predicted class probabilities.

\paragraph{Implementation Details.}
The Swin Transformer used in the first stage was pre-trained on ImageNet~\cite{DeDo09}, and zero padding of 37 pixels was applied to reduce border artifacts.
The patch operator $\Patch$ subdivides each image into overlapping patches of size $224\times224$ with a stride of 150 pixels.
For TTA, we used all rotation angles, i.e., $N_r = 360$.
The model was trained for 50 epochs with a batch size of 4.
Data augmentation included vertical and horizontal flips applied to the whole slide, each applied with probability 0.5.
Individual patches underwent random rotations of $0^\circ$, $30^\circ$, $60^\circ$, $90^\circ$, $120^\circ$, or $150^\circ$.
Gaussian blur was applied with a kernel size of 11 and $\sigma \in [0.05, 0.1]$, along with brightness jittering with a factor of 0.05.

In the second stage, we employed a U-Net with three down- and upsampling layers as the learned regularizer, using two $7\times7$ convolutional layers at each scale.
The network contains 12, 24, 48, and 96 feature maps across four scales.
For binary tasks (BCSS and RINGS), the network takes a 7-channel input (image, histogram, and 1-channel prediction) with 1,369,573 parameters.
In the in-house 3-class setting, the input increases to 15 channels (image, class histograms, and 3-channel prediction), yielding 1,374,277 parameters.
The network was trained for 30 epochs with a batch size of 1 due to memory constraints.
Data augmentation, including Gaussian blur and brightness jittering, matched the first stage and was applied to the image input, while random vertical and horizontal flips were applied to all channels.

Both stages used the AdamW optimizer~\cite{LoFr19,KiBa15} with a learning rate of $3 \times 10^{-5}$.
The learnable scaling parameters $\tau$ and $\lambda$ were updated using an individual learning rate of 0.05.
The number of discretization steps used during training to estimate $\tau$, $\lambda$, and $\theta$ was set to $S = 20$, with initial values $\tau = \lambda = 1$.

After 30 epochs, $\tau$ and $\lambda$ converged to $14.84$ and $2.56$ on the RINGS dataset and $1.56$ and $10.05$ on the BCSS dataset.
Following the ablation study, the number of bins for the histogram was set to $N_h=3$, with an initial noise strength $\sigma^0 = 0.3$, and the number of discretization steps during inference was set to $S=50$.
Training in the first stage required approximately 16 hours, while the second stage required approximately 50 hours on a single NVIDIA A40 GPU.

\subsection{Comparison with State-of-the-Art Methods}\label{subsec4_2}

\begin{table}[htb]
  \centering
  \small
  \setlength\tabcolsep{1pt}
  \begin{tabular}{@{}lcccccc@{}}
    \toprule
    DS & Type & Method & Dice $\uparrow$ & mIoU $\uparrow$ & 95HD $\downarrow$\\
    \midrule
    {\multirow{20}{*}{\rotatebox[origin=c]{90}{RINGS dataset}}}
    & Unsup. & MedSAM~\cite{MaYu24} & 0.656$\pm 0.21$ & 0.524$\pm 0.23$ & 418$\pm 316$ \\
    \cmidrule{2-6}
    &{\multirow{7}{*}{\rotatebox[origin=c]{90}{MIL}}}
    &  LPLP~\cite{MaSu24} & 0.587$\pm 0.24$ & 0.398$\pm 0.19$ & 321$\pm 262$ \\
    & & SA-MIL~\cite{li23weak} & 0.785$\pm 0.19$ & 0.678$\pm 0.2$ & $271\pm 268$\\
    & & WSSS-T~\cite{han22mul} & 0.789$\pm 0.19$ & 0.674$\pm 0.19$ & 286$\pm 260$ \\
    & & C2C~\cite{shar21c} & 0.781$\pm 0.19$ & 0.671$\pm 0.21$ & 251$\pm 251$ \\
    & & PistoSeg~\cite{fang23weak} & 0.798$\pm 0.19$ & 0.682$\pm 0.2$ & 278$\pm 273$ \\
    & & OEEM~\cite{li22onl} & 0.791$\pm 0.23$ & 0.679$\pm 0.23$ & 313$\pm 287$ \\
    & & URN~\cite{li22urn} & $\mathbf{0.867\pm 0.10}$ & $\mathbf{0.783\pm 0.15}$ & $\mathbf{115\pm 98}$ \\
    \cmidrule{2-6}
    &{\multirow{8}{*}{\rotatebox[origin=c]{90}{LLP}}}
    & SA-MIL~\cite{li23weak} & 0.42$\pm 0.2$ & 0.243$\pm 0.17$ & \ 403$\pm 299$ \\
    & & C2C~\cite{shar21c} & 0.367$\pm 0.11$ & 0.21$\pm 0.13$ & \ 370$\pm 294$ \\
    & & FGNet~\cite{zh21joint} & 0.788$\pm 0.21$ & 0.683$\pm 0.22$ & \ 282$\pm 292$ \\
    & & FlipCAM~\cite{24flipcam} & 0.5821$\pm 0.23$ & 0.4106$\pm 0.22$ & 432$\pm 356$ \\
    & & 1st stage $\bar x$ & 0.775$\pm 0.2$ & 0.648$\pm 0.21$ & \ 265$\pm 230$ \\
    & & VSLP$_{ULA}$ & \textcolor[rgb]{0,0,1}{0.829$\pm 0.21$} & 0.687$\pm 0.22$ & 576$\pm 368$ \\
    & & VSLP$_{GMM}$ & 0.801$\pm 0.27$ & 0.679$\pm 0.25$ & 251$\pm 232$ \\
    & & VSLP$_{Diff.}$ & 0.819$\pm 0.21$ & \textcolor[rgb]{0,0,1}{0.689$\pm 0.22$} & \textcolor[rgb]{0,0,1}{146$\pm 131$} \\
    \cmidrule{2-6}
    & {\multirow{3}{*}{Full sup.}} & U-Net~\cite{ron15u} & 0.945$\pm 0.03$ & 0.898$\pm 0.05$ & 60$\pm 57.7$ \\
    & & SwinUnet~\cite{22swinUnet} & 0.875$\pm 0.0.12$ & 0.791$\pm 0.16$ & $173\pm 145$\\
    & & TransUnet~\cite{21transunet} & 0.851$\pm 0.17$ & 0.793$\pm 0.17$ & $177\pm 155$\\
    \midrule
    {\multirow{20}{*}{\rotatebox[origin=c]{90}{BCSS dataset}}}
    & Unsup. & MedSAM~\cite{MaYu24} & 0.597$\pm 0.25$ & 0.472$\pm 0.25$ & 536$\pm 318$ \\
    \cmidrule{2-6}
    &{\multirow{7}{*}{\rotatebox[origin=c]{90}{MIL}}}
    & LPLP~\cite{MaSu24} & 0.489$\pm 0.18$ & 0.341$\pm 0.14$ & \ 378$\pm 309$ \\
    & & SA-MIL~\cite{li23weak} & 0.696$\pm 0.26$ & 0.583$\pm 0.26$ & 375$\pm 283$ \\
    & & WSSS-T~\cite{han22mul} & 0.737$\pm 0.21$ & 0.612$\pm 0.22$ & $\mathbf{336\pm 307}$ \\
    & & C2C~\cite{shar21c} & 0.722$\pm 0.26$ & 0.616$\pm 0.25$ & 340$\pm 273$ \\
    & & PistoSeg~\cite{fang23weak} & 0.672$\pm 0.27$ & 0.56$\pm 0.27$ & 418$\pm 318$ \\
    & & OEEM~\cite{li22onl} & 0.662$\pm 0.23$ & 0.533$\pm 0.22$ & 416$\pm 299$ \\
    & & URN~\cite{li22urn} & 0.640$\pm 0.25$ & 0.516$\pm 0.24$ & 339$\pm 255$ \\
    \cmidrule{2-6}
    &{\multirow{8}{*}{\rotatebox[origin=c]{90}{LLP}}}
    & SA-MIL~\cite{li23weak} & 0.532$\pm 0.3$ & 0.42$\pm 0.28$ & \ 493$\pm 345$ \\
    & & C2C~\cite{shar21c} & 0.499$\pm 0.27$ & 0.374$\pm 0.23$ & \ 502$\pm 321$ \\
    & & FGNet~\cite{zh21joint} & 0.576$\pm 0.26$ & 0.447$\pm 0.24$ & \ 487$\pm 319$ \\
    & & FlipCAM~\cite{24flipcam} & 0.479$\pm 0.25$ & 0.315$\pm 0.22$ & 649$\pm 395$ \\
    & & 1st stage $\bar x$ & 0.763$\pm 0.26$ & 0.638$\pm 0.25$ & 358$\pm 307$ \\
    & & VSLP$_{ULA}$ & 0.69$\pm 0.24$ & 0.44$\pm 0.3$ & 749$\pm 356$ \\
    & & VSLP$_{GMM}$ & $\mathbf{0.786\pm 0.25}$ & $\mathbf{0.655\pm 0.29}$ & 350$\pm 343$ \\
    & & VSLP$_{Diff.}$ & \textcolor[rgb]{0,0,1}{$0.779\pm 0.23$} & \textcolor[rgb]{0,0,1}{$0.639\pm 0.28$} & \textcolor[rgb]{0,0,1}{340$\pm 330$} \\
    \cmidrule{2-6}
    & {\multirow{3}{*}{Full sup.}} & U-Net~\cite{ron15u} & 0.838$\pm 0.24$ & 0.750$\pm 0.24$ & 286$\pm 260$ \\
    & & SwinUnet~\cite{22swinUnet} & 0.825$\pm 0.23$ & 0.716$\pm 0.22$ & 275$\pm 235$\\
    & & TransUnet~\cite{21transunet} & 0.816$\pm 0.24$ & 0.684$\pm 0.22$ & $278\pm 231$\\
    \bottomrule
  \end{tabular}
  \caption{Quantitative comparison with unsupervised, MIL, LLP, and fully supervised methods on the RINGS and BCSS datasets. The best results are shown in bold, and the second-best results are highlighted in blue. Note that the 2nd stage consistently improves upon the first stage.}
  \label{tab:comparison_results}
\end{table}

\begin{table}[htb]
\centering
\small
\begin{tabular}{l l c c c}
\toprule
DS  & Method & Dice $\uparrow$ & mIoU $\uparrow$ & 95HD $\downarrow$ \\
\midrule
\multirow{6}{*}{\rotatebox[origin=c]{90}{RINGS-noise}}
& C2C~\cite{shar21c} & 0.324$\pm$0.140 & 0.202$\pm$0.100 & 395$\pm$301 \\
& SA-MIL~\cite{li23weak} & 0.489$\pm$0.231 & 0.354$\pm$0.198 & 397$\pm$300 \\
& 1st stage $\bar x$ & 0.752$\pm$0.222 & 0.642$\pm$0.229 & \textcolor[rgb]{0,0,1}{265$\pm$244} \\
& VSLP$_{ULA}$   & 0.758$\pm$0.123 & 0.617$\pm$0.226 & 579$\pm$342 \\
& VSLP$_{GMM}$   & $\mathbf{0.807\pm0.127}$ & $\mathbf{0.675\pm0.224}$ & 474$\pm$249 \\
& VSLP$_{Diff.}$ & \textcolor[rgb]{0,0,1}{0.794$\pm$0.124} & \textcolor[rgb]{0,0,1}{0.655$\pm$0.225} & $\mathbf{251\pm201}$ \\
\midrule
\multirow{6}{*}{\rotatebox[origin=c]{90}{RINGS-Path.}}
& C2C~\cite{shar21c} & 0.484$\pm$0.227 & 0.396$\pm$0.225 & 435$\pm$343 \\
& SA-MIL~\cite{li23weak} & 0.379$\pm$0.214 & 0.256$\pm$0.175 & 497$\pm$299 \\
& 1st stage $\bar x$ & \textcolor[rgb]{0,0,1}{0.720$\pm$0.198} & \textcolor[rgb]{0,0,1}{0.593$\pm$0.201} & \textcolor[rgb]{0,0,1}{298$\pm$260} \\
& VSLP$_{GMM}$ & 0.587$\pm$0.251 & 0.477$\pm$0.223 & 483$\pm$315 \\
& VSLP$_{ULA}$   & 0.678$\pm$0.168 & 0.569$\pm$0.226 & 565$\pm$247 \\
& VSLP$_{Diff.}$ & $\mathbf{0.749\pm0.211}$ & $\mathbf{0.628\pm0.203}$ & $\mathbf{284\pm200}$ \\
\bottomrule
\end{tabular}
\caption{Performance comparison on the RINGS dataset under label noise conditions. RINGS-noise denotes the dataset with 10\% perturbed global proportions, while RINGS-Path refers to coarse pathologist-provided annotations.}
\label{tab:rings_noise}
\end{table}

\begin{table*}[htb]
  \centering
  \small
  \begin{tabular}{cccccccc}
    \toprule
    DS & Method & Acc.\ $\uparrow$ & Pre.\ $\uparrow$ & Recall $\uparrow$ & F1 $\uparrow$ & MAE $\downarrow$ & MSE $\downarrow$ \\
    \midrule
    {\multirow{6}{*}{\rotatebox[origin=c]{90}{In-house}}}
    & SA-MIL~\cite{li23weak} & 0.667 & 0.442 & 0.612 & 0.512 & 0.213 & 0.077 \\
    & C2C~\cite{shar21c} & 0.429 & 0.143 & 0.333 & 0.20 & 0.341 & 0.221 \\
    & 1st stage $\bar x$ & \textcolor[rgb]{0,0,1}{0.937} & \textcolor[rgb]{0,0,1}{0.947} & 0.928 & 0.934 & 0.103 & \textcolor[rgb]{0,0,1}{0.025} \\
    & VSLP$_{ULA}$ & 0.937 & 0.947 & 0.928 & 0.934 & 0.158 & 0.032 \\
    & VSLP$_{GMM}$ & 0.937 & 0.939 & $\mathbf{0.933}$ & $\mathbf{0.935}$ & $\mathbf{0.095}$ & 0.059 \\
    & VSLP$_{Diff.}$ & $\mathbf{0.937}$ & $\mathbf{0.947}$ & \textcolor[rgb]{0,0,1}{0.928} & \textcolor[rgb]{0,0,1}{0.934} & \textcolor[rgb]{0,0,1}{0.101} & $\mathbf{0.023}$ \\
    \bottomrule
  \end{tabular}
  \caption{Quantitative comparison with LLP methods on our in-house dataset. Best results are shown in bold, and second-best results are highlighted in blue.}
  \label{tab:inhouse_results}
\end{table*}

The segmentation performance of our method was compared with recent MIL methods (SA-MIL~\cite{li23weak}, WSSS-T~\cite{han22mul}, C2C~\cite{shar21c}, PistoSeg~\cite{fang23weak}, OEEM~\cite{li22onl}, URN~\cite{li22urn}), LLP methods (FGNet~\cite{zh21joint}, FlipCAM~\cite{24flipcam}), unsupervised methods (MedSAM~\cite{MaYu24}), and fully supervised methods (U-Net~\cite{ron15u}, SwinUnet~\cite{22swinUnet}, TransUnet~\cite{21transunet}).
Some methods are not originally labeled as MIL or LLP, so we categorize them according to their supervision type to ensure a consistent comparison.
To align with our label format, we adapt SA-MIL~\cite{li23weak} and C2C~\cite{shar21c} by replacing their instance-level MIL supervision with global label proportions, effectively converting them into LLP methods.
Since the LLP phase of LPLP~\cite{MaSu24} is incompatible with binary-class settings, we evaluate only its MIL component.
All methods were trained on the RINGS and BCSS datasets, using their respective open-source code.
On our in-house dataset, which contains only global label proportions, patch-level classification labels required by the MIL stage of LPLP are unavailable.
Therefore, we report results solely for the adapted SA-MIL and C2C methods on this dataset.

\paragraph{Experimental Setting.}
For MIL-based methods, histopathology images are resized to multiples of 128 while preserving their aspect ratios.
For LLP-based methods, full $1500\times1500$ images are processed and split into $128\times128$ patches with a stride of 128; patch outputs are aggregated to obtain image-level category probabilities using the corresponding loss formulations.
MedSAM~\cite{MaYu24} is evaluated using its open-source implementation; as unsupervised methods lack foreground/background labels, we report the better-performing result between binary and inverse mask predictions, following prior work.
Fully supervised baselines are trained using pixel-level annotations under their standard settings.
All methods operate on original-resolution images, with evaluation metrics computed after reconstruction of the full segmentation maps.

\paragraph{Quantitative Results.}

Table~\ref{tab:comparison_results} reports results on the RINGS and BCSS datasets, where the second-stage refinement consistently improves first-stage predictions across all metrics.
LLP-based methods generally underperform MIL approaches because they rely only on global image-level proportions, while MIL benefits from patch-level supervision.
Although URN achieves higher scores on RINGS, it is not directly comparable, as it uses MIL supervision, whereas our method operates under the weaker LLP setting.
Within the LLP regime, our VSLP variants achieve state-of-the-art performance, with VSLP$_{Diff.}$ and VSLP$_{GMM}$ yielding the strongest results.
Our method also shows greater cross-dataset stability on BCSS, highlighting its robustness despite weaker supervision.

Table~\ref{tab:rings_noise} evaluates the robustness of the proposed method to noisy and coarse supervision.
Under 10\% proportion perturbations (RINGS-noise), VSLP$_{GMM}$ achieves the best Dice and mIoU, highlighting its stability when global statistics are imprecise.
With coarse pathologist labels (RINGS-Path), VSLP$_{Diff.}$ outperforms all other methods, indicating superior adaptability to weak and spatially imprecise supervision.
These results show that the proposed refinement framework remains effective across different forms of label noise.

Table~\ref{tab:inhouse_results} presents the performance on our in-house LLP dataset.
We compare the existing methods SA-MIL~\cite{li23weak} and C2C~\cite{shar21c} with our proposed VSLP model.
Notably, our method (both the 1st stage and 2nd stage) substantially outperforms the baselines, with VSLP$_{Diff.}$ achieving the best overall performance.

\paragraph{Qualitative Results.}

\begin{figure*}[htb]
\begin{tikzpicture}
\node (fig) at (-2,0) {\includegraphics[width=0.98\linewidth]{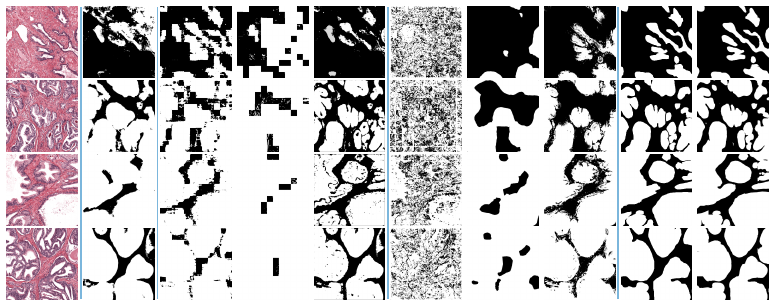}};
\tikzmath{
    \w=1.2;
    \dw=\w+0.6;
    \totalwidth = 9*\dw;
}
\foreach \e [count=\x from 0] in {input, MedSAM\\\cite{MaYu24}, SA-MIL\\\cite{li23weak},WSSS\\\cite{han22mul},PistoSeg\\\cite{fang23weak},SA-MIL\\(LLP)\cite{li23weak},1st stage $\bar x$ (Ours),VSLP$_{Diff.}$ (Ours), ground truth, U-Net\\\cite{ron15u}} {
    \node[below of=fig, yshift=-80, xshift={(\x*\dw - 4.5*\dw)*1cm}] {\footnotesize
    \parbox{1.1cm}{\centering \e}
    };
}
\end{tikzpicture}
\caption{Comparison of our method with the state-of-the-art approaches on samples from the RINGS~\cite{salvi21hybrid} dataset. Note that MedSAM~\cite{MaYu24} is fully unsupervised, and U-Net is fully supervised on the GT segmentation, while all the other methods are MIL and LLP.}
\label{fig:rings_comp}
\end{figure*}

Figure~\ref{fig:rings_comp} shows segmentation comparisons between our method and state-of-the-art approaches on the RINGS dataset.
For each method type, we show the best-performing model.
The purely unsupervised MedSAM performs inconsistently, with some segmentations surpassing those of MIL and LLP methods, while failing in other cases.
SA-MIL demonstrates better localization but tends to produce fragmented or incomplete segmentation, particularly along gland boundaries.
PistoSeg offers improved continuity but still suffers from over-segmentation in complex regions.
The LLP-adapted SA-MIL baseline captures only general textures without accurate structure.
In contrast, our method delivers more coherent and precise segmentations.

\begin{figure*}[htb]
\begin{tikzpicture}
\node (fig) at (-2,0) {\includegraphics[width=0.98\linewidth]{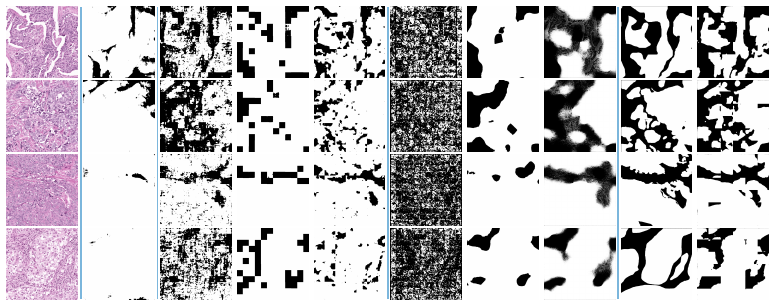}};
\tikzmath{
    \w=1.2;
    \dw=\w+0.6;
    \totalwidth = 9*\dw;
}
\foreach \d/\e [count=\x from 0] in {rgb/input, MEDSAM/MedSAM\\\cite{MaYu24}, sa\_mil/SA-MIL\\\cite{li23weak},WSSS-TISSUE/WSSS\\\cite{han22mul},PistoSeg/PistoSeg\\\cite{fang23weak},sa\_mil\_llp/SA-MIL\\(LLP)\cite{li23weak}, our\_first/1st stage $\bar x$ (Ours),our\_second/VSLP$_{Diff.}$ (Ours), gt/ground truth, unet/U-Net\\\cite{ron15u}} {
    \node[below of=fig, yshift=-80, xshift={(\x*\dw - 4.5*\dw)*1cm}] {\footnotesize
    \parbox{1.1cm}{\centering \e}
    };
}
\end{tikzpicture}
\caption{Comparison of our method with the state-of-the-art approaches on samples from the BCSS~\cite{amgad2019struc} dataset.}
\label{fig:bcss_comp}
\end{figure*}

In most cases on the BCSS dataset (Figure~\ref{fig:bcss_comp}), MedSAM struggles to capture the complex and diverse tumor structures, often missing key regions entirely.
SA-MIL's predictions can appear noisy, while WSSS-T produces coarse, block-like segmentations that fail to capture fine morphological details.
PistoSeg achieves better continuity but still lacks boundary precision.
The LLP-adapted SA-MIL baseline fails to meaningfully capture tumor morphology.
In contrast, our method, particularly in the second stage, refines uncertain regions, producing segmentation maps that visually align more closely with the ground truth.
Nonetheless, the BCSS dataset remains highly challenging due to its variability and fine-grained structures.

\begin{figure*}[htb]
    \centering
    \includegraphics[width=.8\linewidth]{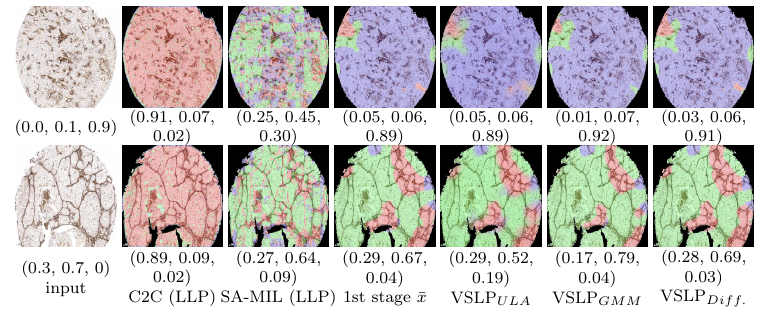}
    \caption{Comparison of our method with LLP-adapted SA-MIL and C2C on the in-house dataset. Red indicates high branching, green indicates low branching, and blue indicates sinusoid. Global class proportions are provided by an experienced pathologist and serve as the ground-truth labels. Predicted class proportions are computed by summing segmented pixels for each class. Notably, all of our models generate meaningful segmentation and accurate class proportions, enhancing trust in the predictions.}
    \label{fig:multi_comp}
\end{figure*}

Figure~\ref{fig:multi_comp} compares segmentation on our in-house dataset, which comprises three vascular patterns: high branching (red), low branching (green), and sinusoid (blue).
An experienced pathologist assigned global label proportions of 0.0, 0.1, and 0.9 for the first example, and 0.3, 0.7, and 0.0 for the second example, corresponding to high branching, low branching, and sinusoid, respectively.
While the LLP-adapted SA-MIL baselines capture only coarse structures and misclassify certain regions, our two-stage approach yields more coherent and morphologically consistent segmentation maps.
Notably, the second stage effectively corrects misclassifications in fine-grained regions.

\subsection{Ablation Study}\label{subsec4_3}
This section analyzes the key factors influencing our method, focusing on the quality of the first-stage posterior and the roles of global proportion constraints and the learned regularizer.
Additional ablation experiments are reported in the supplementary materials.

\paragraph{Influence of 1st Stage Quality.}
The effectiveness of the second stage depends on the quality and stability of the first-stage posterior $\bar{x}$, which are controlled by the patch configuration and the number of TTA rotations.
As shown in Table~\ref{tab:patch_size}, smaller patches with overlapping strides (224, 150) better preserve boundary localization and produce more informative posteriors, whereas larger patches blur boundaries and limit refinement.
Although the second stage improves performance across all patch settings, the gains saturate when the input posterior becomes overly coarse.
Table~\ref{tab:tta_rotation} further shows that TTA stabilizes the posterior, as insufficient rotations often fail to recover fine structures.
Increasing the number of rotations improves uncertainty estimation and correction, with diminishing returns beyond 180--360 rotations.
Overall, the variational refinement is corrective rather than generative, refining stable posteriors but not compensating for degraded first-stage predictions.
Similar behavior across different first-stage backbones indicates that the refinement is not tied to a specific architecture.

\begin{table}[htb]
  \centering
  \small
  \setlength\tabcolsep{7pt}
  \begin{tabular}{@{}ccccccc@{}}
    \toprule
    Method & Patch & Stride & Dice $\uparrow$ & mIoU $\uparrow$ & 95HD $\downarrow$ \\
    \midrule
    \multirow{3}{*}{1st stage $\bar x$} & 224 & 224 & 0.7589 & 0.6341 & $\mathbf{265.0238}$ \\
    & 224 & 150 & $\mathbf{0.7749}$ & $\mathbf{0.6477}$ & 265.3500 \\
    & 384 & 250 & 0.7285 & 0.5986 & 443.6101 \\
    \midrule
    \multirow{3}{*}{VSLP$_{ULA}$} & 224 & 224 & 0.8099 & 0.6587 & 577.1805 \\
    & 224 & 150 & $\mathbf{0.8293}$ & $\mathbf{0.6871}$ & $\mathbf{576.8200}$ \\
    & 384 & 250 & 0.7802 & 0.6175 & 582.7280 \\
    \midrule
    \multirow{3}{*}{VSLP$_{Diff.}$} & 224 & 224 & 0.7712 & 0.6408 & 214.6227 \\
    & 224 & 150 & $\mathbf{0.8195}$ & $\mathbf{0.6894}$ & $\mathbf{145.7532}$ \\
    & 384 & 250 & 0.7936 & 0.6724 & 235.9983 \\
    \midrule
    \multirow{3}{*}{VSLP$_{GMM}$} & 224 & 224 & 0.7930 & 0.6223 & 301.6642 \\
    & 224 & 150 & $\mathbf{0.8013}$ & $\mathbf{0.6792}$ & $\mathbf{250.9732}$ \\
    & 384 & 250 & 0.7428 & 0.5580 & 476.1845 \\
    \bottomrule
  \end{tabular}
  \caption{Effect of patch size and stride on segmentation performance for the first-stage and second-stage VSLP variants.}
  \label{tab:patch_size}
\end{table}

\begin{table}[htb]
  \centering
  \small
  \setlength\tabcolsep{8pt}
  \begin{tabular}{@{}cccccc@{}}
    \toprule
    Method & TTA Rotations & Dice $\uparrow$ & mIoU $\uparrow$ & 95HD $\downarrow$ \\
    \midrule
    \multirow{3}{*}{1st stage $\bar x$}
    & 24 & 0.7681 & 0.6408 & 352.4732 \\
    & 180 & $\mathbf{0.7749}$ & $\mathbf{0.6478}$ & 347.7140 \\
    & 360 & 0.7749 & 0.6477 & $\mathbf{265.3500}$ \\
    \midrule
    \multirow{3}{*}{VSLP$_{ULA}$}
    & 24 & 0.7701 & 0.6672 & $\mathbf{331.3889}$ \\
    & 180 & 0.8204 & 0.6856 & 483.4881 \\
    & 360 & $\mathbf{0.8293}$ & $\mathbf{0.6871}$ & 576.8200 \\
    \midrule
    \multirow{3}{*}{VSLP$_{Diff.}$}
    & 24 & 0.7665 & 0.6414 & 347.6053 \\
    & 180 & $\mathbf{0.8310}$ & 0.6568 & 202.2721 \\
    & 360 & 0.8195 & $\mathbf{0.6894}$ & $\mathbf{145.7532}$ \\
    \midrule
    \multirow{3}{*}{VSLP$_{GMM}$}
    & 24 & 0.7664 & 0.6156 & 579.8176 \\
    & 180 & 0.7876 & 0.6440 & 315.8539 \\
    & 360 & $\mathbf{0.8013}$ & $\mathbf{0.6792}$ & $\mathbf{250.9732}$ \\
    \midrule
    Method & Model & Dice $\uparrow$ & mIoU $\uparrow$ & 95HD $\downarrow$ \\
    \midrule
    \multirow{3}{*}{1st stage $\bar x$}
    & SwinUnet & $\mathbf{0.7817}$ & $\mathbf{0.6856}$ & 323.4943\\
    & TransUnet & 0.6719 & 0.5490 & 441.5711 \\
    & Swin-T & 0.7749 & 0.6477 & $\mathbf{265.3500}$ \\
    \bottomrule
  \end{tabular}
  \caption{Effect of TTA rotation count on segmentation performance. The upper part reports performance as a function of the number of TTA rotations for different second-stage variants, using the best-performing first-stage patch size (224, 150). The lower part compares different first-stage backbone models under the same TTA setting.}
  \label{tab:tta_rotation}
\end{table}

\paragraph{Effect of Proportions and Regularizer.}

We next evaluate the contributions of global proportion constraints and the learned regularizer (Table~\ref{tab:vslp_stats}).
High-contrast slides have class proportion differences greater than 0.3, while low-contrast slides exhibit more ambiguous boundaries.
Across both subsets, VSLP$_{Diff.}$ and VSLP$_{GMM}$ provide the strongest refinements, whereas VSLP$_{ULA}$ performs best in low-contrast cases where first-stage predictions are less reliable.
Replacing true proportions with random values causes a substantial performance drop, confirming the importance of accurate global statistics.
Removing the learned regularizer yields fragmented or degenerate outputs, showing that Wasserstein fidelity alone cannot enforce spatial coherence.
Overall, the global proportions control class ratios, while the regularizer governs spatial structure, and both are essential for robust performance across contrast levels.

\begin{table}[htb]
\centering
\small
\setlength\tabcolsep{6pt}
\begin{tabular}{@{}ccccc@{}}
\toprule
Data type & Method & Dice $\uparrow$ & mIoU $\uparrow$ & 95HD $\downarrow$ \\
\midrule
\multirow{4}{*}{High contrast}
  & 1st stage $\bar x$ & 0.7385 & 0.6363 & 246 \\
  & VSLP$_{ULA}$ & 0.7307 & 0.6345 & 423 \\
  & VSLP$_{GMM}$ & 0.7659 & $\mathbf{0.6535}$ & 245 \\
  & VSLP$_{Diff.}$& $\mathbf{0.7673}$ & 0.6440 & $\mathbf{163}$ \\
\midrule
\multirow{4}{*}{Low contrast}
  & 1st stage $\bar x$ & 0.8244 & 0.7105 & 114 \\
  & VSLP$_{ULA}$ & $\mathbf{0.8716}$ & $\mathbf{0.7800}$ & 175 \\
  & VSLP$_{GMM}$ & 0.8528 & 0.7185 & 116 \\
  & VSLP$_{Diff.}$ & 0.8513 & 0.7530 & $\mathbf{64}$ \\
\midrule
\multirow{4}{*}{Random Proportions}
  & 1st stage $\bar{x}$ & 0.060 & 0.060 & inf \\
  & VSLP$_{ULA}$ & 0.3983 & 0.2639 & 579 \\
  & VSLP$_{GMM}$ & 0.5055 & 0.3753 & $\mathbf{579}$ \\
  & VSLP$_{Diff.}$& $\mathbf{0.5518}$ & $\mathbf{0.4540}$ & 620 \\
\midrule
\multirow{3}{*}{No regularizer}
  & VSLP$_{ULA}$ & 0.3041 & 0.1793 & 495 \\
  & VSLP$_{GMM}$ & 0.0600 & 0.0600 & inf \\
  & VSLP$_{Diff.}$& $\mathbf{0.5449}$ & $\mathbf{0.3745}$ & $\mathbf{437}$ \\
\bottomrule
\end{tabular}
\caption{Performance of VSLP variants under different data conditions.}
\label{tab:vslp_stats}
\end{table}

\subsection{Limitations of Our Approach}\label{subsec4_4}

We assume that segmentation masks follow regular spatial structures encouraged by the learned prior.
Objects that deviate strongly from the training shape distribution may degrade second-stage performance.
Our single-component GMM variant assumes unimodal, symmetric pixel-wise uncertainty and therefore cannot model more complex distributions.
This limitation could be addressed by employing multi-component Gaussian mixtures and sharing parameters across neighboring pixels to capture uncertainty and spatial coherence better.
Since no explicit boundary refinement or post-processing is applied, segmentations may exhibit locally jagged edges, which could be mitigated with lightweight post-processing without altering the core formulation.
Finally, differences in staining protocols, such as H\&E versus CD31, introduce domain shifts that may affect model behavior.
Cross-stain generalization is not addressed and would likely require stain normalization or domain transfer as preprocessing~\cite{gan24}.

\section{Conclusion}\label{sec5}

In this work, we propose a novel probabilistic framework for learning from global label proportions.
Our approach treats the output of TTA applied to a fine-tuned Swin Transformer as samples from an underlying probability distribution, which is approximated using pixel-wise histograms.
An iterative optimization scheme incorporating a Wasserstein-based data fidelity term and a learned regularizer then refines the segmentation.
In this way, global proportion supervision is combined with variational refinement to address the inherent ambiguity of segmentation from label proportions.
Evaluation on the RINGS and BCSS datasets demonstrates that our method outperforms state-of-the-art weakly supervised and unsupervised approaches.
Extensive ablation studies highlight the contribution of each stage.
The robustness of our approach in a multiclass setting under label noise is shown on our in-house dataset.
Nevertheless, we aim to extend this approach to model predictive uncertainty by incorporating pixel-wise output distributions, which enable direct visualization of uncertainty in predictions.
This opens the door to uncertainty-aware segmentation, which is particularly valuable in applications such as medical diagnosis.


\bibliography{references}

\end{document}